\documentclass[aps,prl,preprint,groupedaddress,footinbib]{revtex4-1}
\usepackage[dvips]{graphicx}
\begin{document}


\title{Superconducting fluctuations in FeSe$_{0.5}$Te$_{0.5}$ thin films probed via microwave spectroscopy}


\author{Fuyuki Nabeshima}
\email[]{nabeshijma@maeda1.c.u-tokyo.ac.jp}
\affiliation{Department of Basic Science, the University of Tokyo, Meguro, Tokyo 153-8902, Japan}

\author{Kosuke Nagasawa}
\affiliation{Department of Basic Science, the University of Tokyo, Meguro, Tokyo 153-8902, Japan}

\author{Yoshinori Imai}
\affiliation{Department of Physics, Tohoku University, Sendai 980-8578, Japan}

\author{Atsutaka Maeda}
\affiliation{Department of Basic Science, the University of Tokyo, Meguro, Tokyo 153-8902, Japan}


\date{\today}

\begin{abstract}
We investigated the microwave conductivity spectrum of FeSe$_{0.5}$Te$_{0.5}$ epitaxial films on CaF$_2$ in the vicinity of the superconducting transition.
We observed the critical behavior of the superconducting fluctuations in these films with a dimensional crossover from two-dimensional to three-dimensional as the film thickness increased.
From the temperature dependence of the scaling parameters we conclude that the universality class of the superconducting transition in FeSe$_{0.5}$Te$_{0.5}$ is that of the 3D-XY model.
The lower limit of the onset temperature of the superconducting fluctuations, $T^{\mathrm {onset}}$, determined by our measurements was 1.1 $T_{\mathrm c} $, suggesting that the superconducting fluctuations of FeSe$_{0.5}$Te$_{0.5}$ are at least as large as those of optimally- and over-doped cuprates.
\end{abstract}


\maketitle

Iron chalcogenide superconductor, FeSe$_{1-x}$Te$_{x}$, has the largest ratio of the superconducting gap to the Fermi energy, $\Delta / E_{\mathrm F}$, among known superconductors and is expected to be in the crossover region between Bardeen-Cooper-Schrieffer (BCS) and Bose-Einstein condensate (BEC) limits \cite{NatPhys.8.309,SciRep.4.4109,PNAS.111.16309}. 
Therefore FeSe$_{1-x}$Te$_{x}$ attracts much attention as the most suitable material to experimentally investigate possible novel features related to the BCS-BEC crossover in a superconductor.
In the BCS-BEC crossover region, pairs of electrons are formed at temperatures higher than the superconducting transition temperature, $T_{\mathrm c} $, where the bound pairs condensate, and thus, large superconducting fluctuations are expected to be observed. 
A measurement of diamagnetism in FeSe reported that the diamagnetic signal survived up to 20 K\cite{NatCom.7.12843}, which was approximately twice higher than $T_{\mathrm c} $.
However, no other measurement on the superconducting fluctuations has not been reported so far on this material.
Therefore, to elucidate the detailed nature of superconductivity in iron chalcogenides, studies on the superconductivity fluctuations by other techniques are urgently needed.
When the superconducting fluctuations are large, we expect the critical behavior of the fluctuations to be observed.
Most generally, the critical fluctuations of the superconducting transition is described by the XY model since the superconducting order parameter consists of two components (the magnitude and the phase).
Indeed, the XY behavior was observed by high frequency measurements\cite{PRB.73.092504,Ohashi09} in high-$T_{\mathrm c}$ cuprate superconductors, which are another candidate for superconductors in the BCS-BEC crossover regime\cite{LTPhys.32.406,PhysRevB.66.024510,PhysRevB.53.R11961}.
On the other hand, rather different behaviors were reported in measurements of the Nernst effect\cite{Nature.406.486,PRB.64.224519,PRB.73.024510} and the diamagnetism\cite{PRL.95.247002,PRB.81.054510}.
The reason for the discrepancies among these different measurement techniques remains unexplained.
Thus, it is crucially important to investigate the superconducting fluctuations in iron chalcogenides by high-frequency conductivity measurements.

The study on the superconducting fluctuations in iron chalcogenides is important also from another point of view.
Iron chalcogenides are multiband/multigap superconductors.
The critical fluctuations in multigap superconductors have not been investigated quantitatively almost at all.
Some papers reported on the superconducting fluctuations by the scaling analyses of the dc resistivity in magnetic fields and the $I$-$V$ characteristics in FeSe$_{1-x}$Te$_{x}$\cite{PRB.84.174517,JPCM.26.455701} and in other multigap superconductors, such as MgB$_2$\cite{SSC.121.575,PRB.67.020507}.
However, the scaling analysis of dc $I$-$V$ data usually assumes a particular dimension in order to determine the scaling parameters, leading to less convincing analyzed results.
Contrary to this, the frequency-dependent complex conductivity, $\sigma (\omega)$, can provide the unique determination of the scaling parameters without any assumptions\cite{KitanoRSI08}, and thus, is a very effective probe for investigating the superconducting fluctuations.

In this letter we report on the detailed microwave conductivity measurements in FeSe$_{0.5}$Te$_{0.5}$ epitaxial films in the vicinity of $T_{\mathrm c} $.
We did observe the critical behaviors of the superconducting fluctuations.
The temperature dependence of the critical exponents suggests that the superconducting transition is described by the 3D-XY model.
We estimated the lower limit of the onset temperature of the superconducting fluctuations to be $\sim$ 1.1 $T_{\mathrm c} $, which is rather high when compared with conventional superconductors and almost the same as those of optimally- and over-doped cuprate superconductors.

\begin{figure*}[htb]
\includegraphics[width=\linewidth, bb=0 0 375 198]{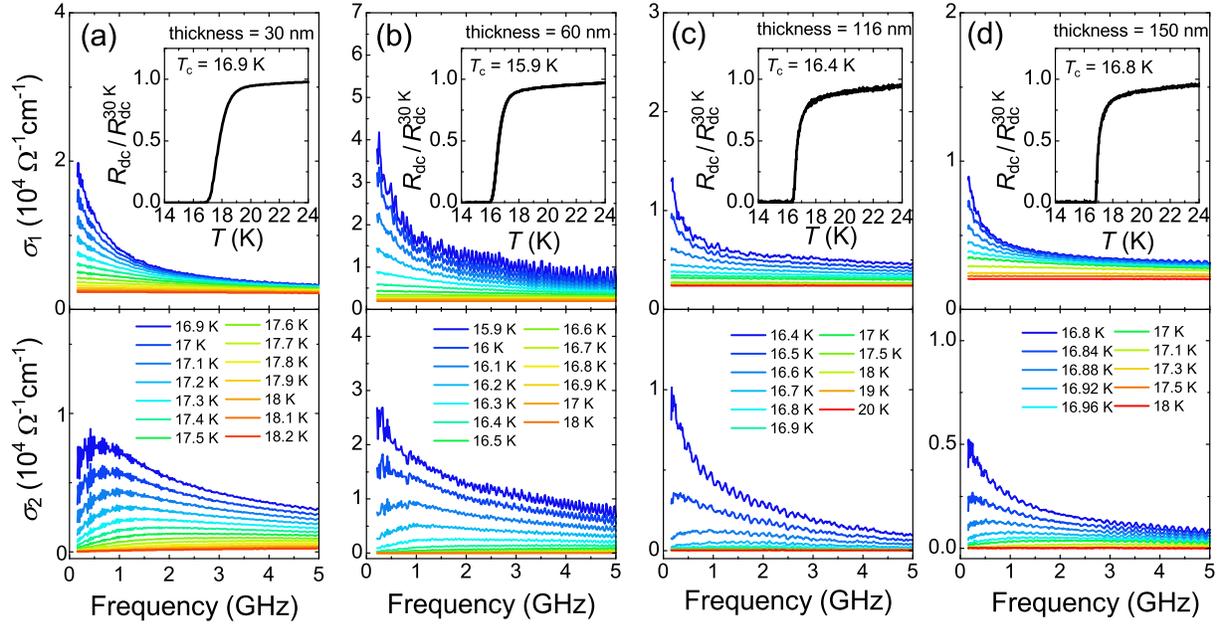}%
\caption{Frequency dependence of the microwave complex conductivity of FeSe$_{0.5}$Te$_{0.5}$ with thickness of (a) 30 nm, (b) 60 nm, (c) 116 nm, and (d) 150 nm. Inset shows the temperature dependence of the normalized resistivity of the films.}
\label{G-sgF}
\end{figure*}

FeSe$_{0.5}$Te$_{0.5}$ epitaxial films were grown on CaF$_2$ by a pulsed laser deposition method using a KrF laser\cite{Imai09,Imai10}.
FeSe$_{0.5}$Te$_{0.5}$ polycrystalline pellets were used as targets. The substrate temperature, the laser repetition rate, and the base pressure were $\sim$ 300$^\circ$C, 20 Hz, and 10$^{-8}$ Torr, respectively.
The thickness of the grown films was measured using a Dektak 6 M stylus profiler.
The grown films show $T_{\mathrm c} $ values larger than bulk values of 15 K (Fig. \ref{G-sgF} inset), which is due to the in-plane compressive lattice strain\cite{Tsukada11,Nabe13}. 
In this paper, $T_{\mathrm c}$ was defined as the temperature where dc resistivity drops to zero.

We measured the complex reflection coefficient, $S_{11}$, of the film placed at the end of the transmission line\cite{KitanoRSI08,PRB77.214517,PRB84.024511}.
Since the measured films are thinner than their skin depth ($\sim 20\ \mu$m at 1 GHz), the complex conductivity, $\sigma = \sigma_1 + i \sigma_2$, can be obtained as follows:
\begin{equation}
\sigma = \frac{1}{tZ_0}\frac{1-S_{11}}{1+S_{11}} , \label{eq:s11}
\end{equation}
where $t$ is the film thickness and $Z_0 = 377\ \Omega$ is the impedance of free space.　

The gold electrodes with the so-called Corbino-disk geometry were sputtered on the film surface. 
The film was connected to a coaxial cable through a modified 2.4 mm jack-to-jack adapter. 
The other end of the transmission line was connected to a vector network analyzer (HP8510C) to measure $S_{11}$ at frequencies from 45 MHz to 10 GHz. 

The experimentally measured reflection coefficient includes the extrinsic attenuation, the reflection, and the phase shift due to the transmission line, etc. 
Thus, the measured reflection coefficient $S_{11}^{\mathrm m}$ can be expressed as follows:
\begin{equation}
S_{11}^{\mathrm m} = E_{\mathrm D} + \frac{E_{\mathrm R}S_{11}}{1-E_{\mathrm S}S_{11}} ,
\end{equation}
where $E_{\mathrm D}$, $E_{\mathrm R}$, and $E_{\mathrm S}$ are complex error coefficients representing the directivity, the reflection tracking, and the source mismatch, respectively.
A set of three independent measurements using samples with the known impedance (or $S_{11}$)  is needed to determine the three error coefficients.
We used a gold film as a short standard, a teflon sheet as an open standard, and FeSe$_{0.5}$Te$_{0.5}$ films in the normal state as a load standard, assuming that the normal state conductivity of FeSe$_{0.5}$Te$_{0.5}$ films was regarded as that in the Hagen-Rubens limit of the Drude conductivity in the measured frequency region\cite{KitanoRSI08}.

Figures \ref{G-sgF} (a)-(d) show the frequency dependence of the complex conductivity in FeSe$_{0.5}$Te$_{0.5}$ films with different thicknesses at several temperatures near $T_{\mathrm c}$.
As the temperature approaches $T_{\mathrm c} $ from above, both $\sigma_1$ and $\sigma_2$ showed a tendency to diverge in the low frequency limit.
This suggests that the contribution of the superconducting fluctuations to $\sigma(\omega)$ was evident with decreasing temperature.

We analyzed the fluctuation conductivity, $\sigma_{\mathrm {fl}}$, at $T \gtrsim T_{\mathrm c}$ in detail. 
First, we subtracted the normal-state conductivity, $\sigma_{\mathrm {n}}$, from the total measured conductivity to extract the fluctuation contribution.
We evaluated $\sigma_{\mathrm {n}}$ as extrapolated values of $\sigma_{\mathrm {dc}} (T)$ from well above $T_{\mathrm c}$. 
Fisher, Fisher, and Huse formulated a dynamic scaling rule on fluctuation conductivity in the vicinity of a superconducting transition, as follows\cite{PRB.43.130}:
\begin{equation}
\sigma_{\mathrm {fl}} (\omega) = \sigma_0 S \left( \frac{\omega}{\omega_0} \right) ,
\end{equation}
where $S$ is a complex universal scaling function, and $\sigma_0$ and $\omega_0$ are scaling parameters.
The temperature dependences of $\sigma_0$ and $\omega_0$ are related to that of a correlation length, $\xi$, which diverges at $T_{\mathrm c} $, as follows:
\begin{eqnarray}
\omega_0 \propto \xi ^{-z} \propto |T/T_{\mathrm c} - 1|^{\nu z} , \nonumber \\
\sigma_0 \propto \xi ^{z+2-d} \propto |T/T_{\mathrm c} - 1|^{-\nu (z+2-d)}, \label{eq:scalingpara}
\end{eqnarray}
where $z$ and $\nu$ are a dynamic and static critical exponent, and $d$ is an effective spatial dimension. 
One of the practical merits of a dynamical scaling analysis with the frequency-dependent complex conductivity is the unique determination of the two scaling parameters. This enables us to evaluate the critical exponents without any assumption, which is in contrast to the case of the $I-V$ characteristic measurements, where we should assume one of three critical exponents (for instance, the dimension) before proceeding the analysis.

\begin{figure*}[htb]
\includegraphics[width=\linewidth, bb=0 0 382 181]{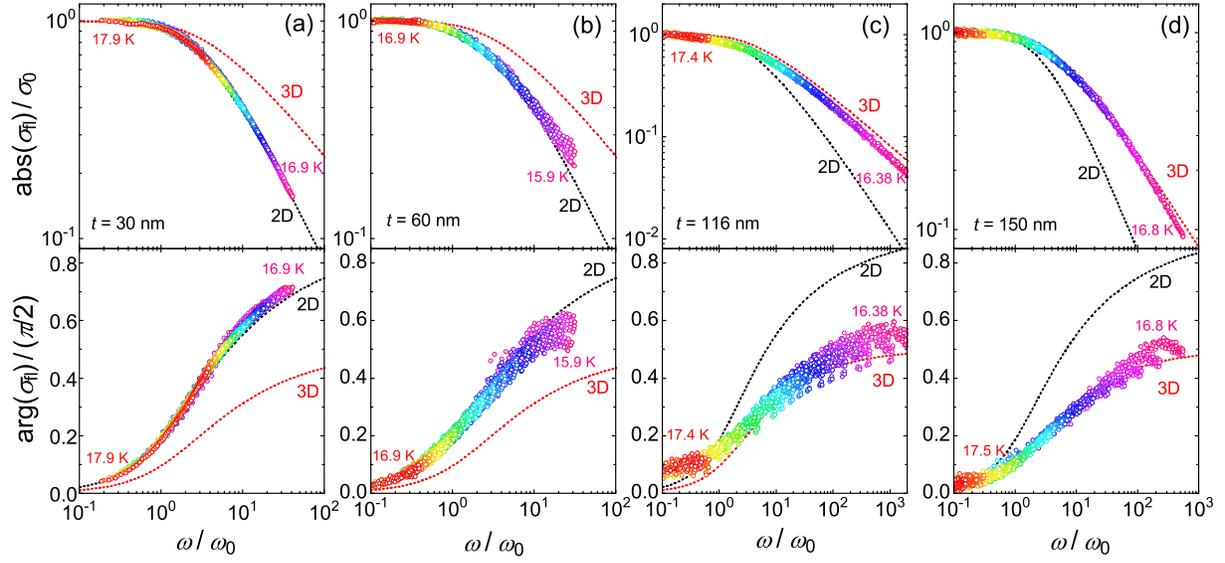}%
\caption{Results of the dynamical scaling analysis for FeSe$_{0.5}$Te$_{0.5}$ thin films with thickness of (a) 30 nm, (b) 60 nm, (c) 116 nm, and (d) 150 nm. Broken curves are those of 2D and 3D fluctuations calculated by Schmidt\cite{Schmidt68}.}
\label{G-Scaling}
\end{figure*}

Figure \ref{G-Scaling} shows the results of the scaling analysis for the films with various film thickness, using the data in the frequency range from 0.3 GHz to 6 GHz for the 30-nm-thick film, 0.3-3 GHz for the 60-nm-thick film, and 0.2-2 GHz for 116- and 150-nm-thick films.
As for the 30-nm-thick sample, each of the magnitude and the phase of the fluctuation conductivity, $|\sigma_{\mathrm {fl}}|$ and $\phi_{\sigma_{\mathrm {fl}}}$, is scaled by one curve, and both curves are almost identical with those expected for two-dimensional fluctuations. 
The similar results were obtained for the 60-nm-thick film. 
Contrary to these, the scaling analysis is not completely satisfactory for the films with thickness of 116 nm and 150 nm.
Figures \ref{G-Scaling} (c) and (d) show the results of the scaling analysis for 116- and 150-nm-thick films, where $\phi_{\sigma_{\mathrm {fl}}}$ is suppressed with increasing frequency. 
The resultant curves were similar to that for three-dimensional fluctuations. 

Our observation is that the superconducting fluctuations are two-dimensional for thin samples and become more three-dimensional as the samples become thicker.
This is a typical behavior of the finite size effect; that is, the two-dimensional behaviors are caused because the out-of-plane coherence length, $\xi_\perp$, becomes much longer than the film thickness at the measured temperatures.
Indeed, these behaviors of the thickness dependence are consistent with $\xi_\perp$ data obtained from transport measurements\cite{ys16jpsj} \footnote{According to ref.\cite{ys16jpsj},  $\xi_\perp \sim 1$ nm in the zero temperature limit. 
Then, we obtain $\xi (T) = \xi (0~\mathrm K) \times (T/T_{\mathrm c} - 1)^{-\nu} \sim 100$ nm ($\nu = 0.67$ for the 3D-XY model) at temperature of $(T/T_{\mathrm c} - 1) \sim 10^{-3}$.
This is consistent with our results of the scaling analysis shown in Fig. \ref{G-Scaling}, that the 30-nm and 60-nm samples were 2D and the 116-nm and 150-nm samples were 3D at the temperature.}
In addition, the similar behavior that $\phi_{\sigma_{\mathrm {fl}}}$ is suppressed at high frequencies was observed in NbN thin films with the thickness of 300 and 450 nm, where a dimensional crossover of the superconducting fluctuations from 3D to 2D was observed\cite{PRB.73.174522}.

\begin{figure}
\includegraphics[width=\linewidth, bb=0 0 585 606]{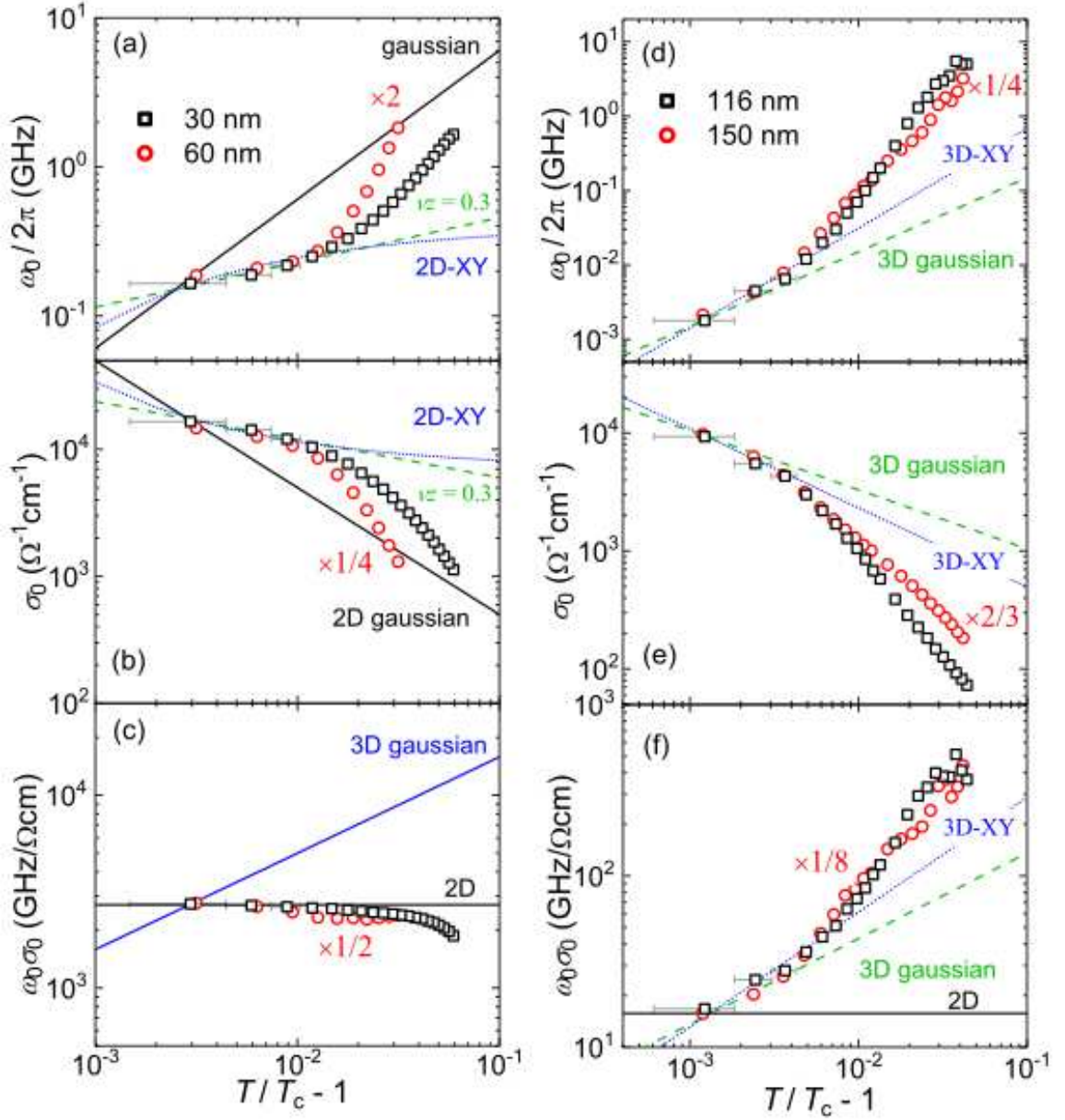}%
\caption{Temperature dependence of the scaling parameters, $\omega_0$, $\sigma_0$, and their product, $\omega_0 \sigma_0$, for FeSe$_{0.5}$Te$_{0.5}$ thin films. The left panel shows those for 30- and 60-nm-thick films, and the right panel shows those for 116- and 150-nm-thick films. Solid and broken lines are theoretically expected behaviors with the corresponding notes.}
\label{G-ScalingPara}
\end{figure}

Figures \ref{G-ScalingPara}(a)-(c) show the temperature dependence of the scaling parameters $\omega_0$, $\sigma_0$ and their product for the 30- and 60-nm-thick films.
Before the quantitative estimation of the critical exponents, it should be noted that the product, $\omega_0 \sigma_0$, showed little temperature dependence (Fig. \ref{G-ScalingPara}(c)). 
This strongly suggests $d = 2$ according to Eq. (\ref{eq:scalingpara}), which is consistent with the results of the scaling curves (Figs \ref{G-Scaling}(a), (b)).
Then, taking a look at $\omega_0$ and $\sigma_0$, their temperature dependence did not follow the theoretically expected behaviors for the 2D gaussian fluctuations.
We obtained the product of the critical exponents $\nu z = $ 0.3 assuming the power-law temperature dependence of $\omega_0$ and $\sigma_0$.
This means the observation of the critical behavior of the superconducting fluctuations.

A two-dimensional superconductor in which the superconducting transition is described by the XY model should show the Berezinskii-Kosterlitz-Thouless (BKT) transition at $T_{\mathrm c} $ \cite{JETP.32.493,KT1973}, where characteristic behaviors are observed such as an exponential temperature dependence of the BKT correlation length in a critical region and the universal jump at $T_{\mathrm c}$\cite{Ohashi09}. 
In the present case, the temperature dependence of $\omega_0$ and $\sigma_0$ can be expressed by an exponential temperature dependence at temperatures very close to $T_{\mathrm c} $. 
Thus, the observed non-gaussian behavior of the relatively thin samples may possibly be the BKT transition, although the universal jump was not observed in the films, which may be due to high frequencies of the measurements.

Next, we discuss the scaling parameters of the thicker films.
Although the scaling analysis is not completely satisfactory for the thicker samples as was shown in Figs. \ref{G-Scaling} (c) and (d), discussion on the scaling parameters of those samples does make sense.
Figs. \ref{G-ScalingPara}(d)-(f) show the temperature dependence of $\omega_0$ and $\sigma_0$ for the 116- and 150-nm-thick films. 
The product, $\omega_0 \sigma_0$, increases with increasing temperature, consistent with the non-2D scaling curves. 
When we evaluate the critical exponents from the temperature dependence of $\omega_0$ and $\sigma_0$, we should keep in mind that the ambiguity in the choice of $T_{\mathrm c} $ largely affects the critical exponents. 
From the data in Figs. \ref{G-ScalingPara}(d)-(f) alone, we cannot distinguish between 3D-XY (blue dotted lines) and 3D gaussian (green broken lines) fluctuations because of the ambiguity in determining $T_{\mathrm c}$, shown by error bars along the horizontal axis in Fig.\ref{G-ScalingPara} (d)-(f).
Nevertheless, 3D gaussian can be excluded because of the following reasons.
When the true fluctuations are 3D gaussian, the temperature dependence of the scaling parameters for a thin sample which shows the 2D fluctuations due to the size effect follows that of the gaussian fluctuations\cite{PRB.73.174522}.
In the present case of iron chalcogenides, however, the non-gaussian behaviors in fluctuations were observed in the thinner film. 
This definitely shows that the true features of the fluctuations of thick films are not 3D gaussian.
Therefore, we conclude that the superconducting fluctuations in FeSe$_{0.5}$Te$_{0.5}$ is originally described by the 3D-XY model.

The observation of the 3D-XY fluctuations in FeSe$_{0.5}$Te$_{0.5}$, which has the stacking structure of two-dimensional layers, is noteworthy.
Indeed, in another layered superconductors, high-$T_{\mathrm c}$ cuprates, the superconducting fluctuations are basically 2D, and become 3D-XY only near the optimal doping, which is understood to be due to the quantum fluctuations\cite{Ohashi09}.
Whether or not the 3D-XY behavior observed in FeSe$_{0.5}$Te$_{0.5}$ is explained by the same mechanism is closely related to the mechanism of the superconductivity in iron chalcogenides.
To clarify this, investigations of the superconducting fluctuations in other compositions are indispensable.
Although bulk crystals of FeSe$_{1-x}$Te$_{x}$ with $0.1< x < 0.4$  cannot be obtained due to the phase separation, the whole compositions are available in thin films\cite{yi15pnas,yiSciRep17}.
Thus, the similar measurements for all of these materials in thin film samples are now underway.

\begin{figure}
\includegraphics[width=\linewidth, bb=0 0 374 310]{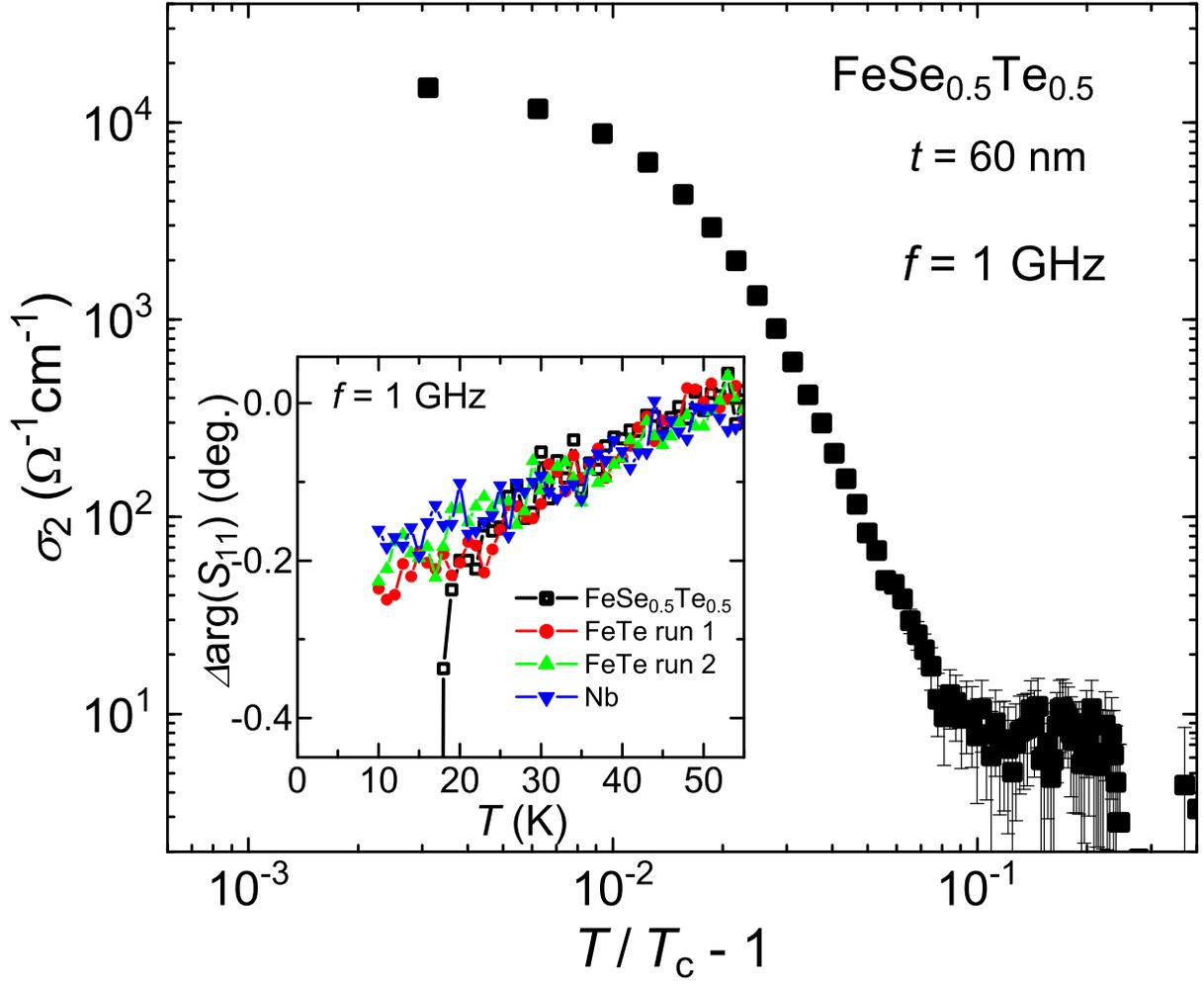}%
\caption{Temperature dependence of the imaginary part of the complex conductivity, $\sigma_2$, at 1 GHz in an FeSe$_{0.5}$Te$_{0.5}$ thin film with thickness of 60 nm. Inset shows the relative temperature variation of the phase of the complex reflection coefficient $S_{11}$ for the FeSe$_{0.5}$Te$_{0.5}$ film. Those of normal metals are also plotted. }
\label{G-Phase}
\end{figure}

Finally, we discuss the onset temperature of the superconductivity, $T^{\mathrm {onset}}$, in FeSe$_{0.5}$Te$_{0.5}$ films.
The response of the superfluid appears in the imaginary part of the complex conductivity, $\sigma_2$, whose variation is sensitively measured by the change in the phase of the complex reflection coefficient, arg($S_{11}$), according to Eq. (\ref{eq:s11}).
However arg($S_{11}$) also changes with temperature due to the thermal expansion of the coaxial cable and other causes.
The inset of Fig. \ref{G-Phase} shows the relative temperature variation of arg($S_{11}$) for the 60-nm-thick FeSe$_{0.5}$Te$_{0.5}$ film together with those of metal films (FeTe and Nb in the normal state). 
arg($S_{11}$)  of the FeSe$_{0.5}$Te$_{0.5}$ film is identical to those for the normal metals above 20 K within the measurement error range, suggesting $T^{\mathrm {onset}}$ is less than 20 K.
Taking the background temperature variation into consideration, we calibrated $S_{11}$ \footnote{We took the temperature-dependent $S_{11}$ of an FeTe film as load data. In addition, to reduce the influence of errors in the phase of $S_{11}$ between the $sample$ and the $reference$ measurements (typically 0.1 deg. at 1 GHz), which is due to the possible difference in the strength of contact between the film and the coaxial cable, we added a constant value to the phase of $S_{11}(T)$ of the load reference so that the phase of $S_{11}^{\mathrm {load}}$ at 50 K to be that of superconducting sample at the same temperature. Then, we calibrate $S_{11}$ of the sample using these data as well as those of Au (Short) and Teflon (Open).}
 and calculated $\sigma_2$ (Figure \ref{G-Phase}).
$\sigma_2$ drops to the measurement error range of $\sim$ 10 $\Omega ^{-1}$cm$^{-1}$  at $T \sim 1.1 T_{\mathrm c}$, which can be regarded as the minimum value of $T^{\mathrm {onset}}$. 
Even the minimum possibility of $T^{\mathrm {onset}} \sim 1.1 T_{\mathrm c}$ is much larger than conventional superconductors, as large as those of optimally- and over-doped cuprates\cite{Ohashi09, Nakamura12}.
This value of $T^{\mathrm {onset}}$, on the other hand, is much smaller than that of FeSe reported in the magnetization measurement\cite{NatCom.7.12843}.
We cannot further discuss the origin of the difference: whether it is due to the difference in compositions, or is due to the difference in measurement techniques.
Thus, again, measurements on samples with other compositions are of great interest.

In conclusion, we measured the detailed microwave conductivity spectrum of FeSe$_{0.5}$Te$_{0.5}$ thin films on CaF$_2$ substrates in the vicinity of the superconducting transition.
We did observe the critical behavior of the superconducting fluctuations. 
Dynamical scaling analyses revealed a dimensional crossover of the superconducting fluctuations from 2D to 3D as the film thickness increases.
From the temperature dependence of the scaling parameters we conclude that the superconducting transition in FeSe$_{0.5}$Te$_{0.5}$ is originally described by the 3D-XY model.
The lower limit of $T^{\mathrm {onset}}$ determined by our measurements was 1.1 $T_{\mathrm c} $, in consistent with large superconducting fluctuations.

\begin{acknowledgments}
We would like to thank M. Hanawa at Central Research Institute of Electric Power Industry for his support in the thickness measurement.
We also thank H. Kitano at Aoyama Gakuin University for fruitful discussion.
This research was supported by JSPS KAKENHI Grant Numbers 15K17697, 14J09315. 
\end{acknowledgments}

\clearpage

\clearpage

\begin{thebibliography}{36}%
\makeatletter
\providecommand \@ifxundefined [1]{%
 \@ifx{#1\undefined}
}%
\providecommand \@ifnum [1]{%
 \ifnum #1\expandafter \@firstoftwo
 \else \expandafter \@secondoftwo
 \fi
}%
\providecommand \@ifx [1]{%
 \ifx #1\expandafter \@firstoftwo
 \else \expandafter \@secondoftwo
 \fi
}%
\providecommand \natexlab [1]{#1}%
\providecommand \enquote  [1]{``#1''}%
\providecommand \bibnamefont  [1]{#1}%
\providecommand \bibfnamefont [1]{#1}%
\providecommand \citenamefont [1]{#1}%
\providecommand \href@noop [0]{\@secondoftwo}%
\providecommand \href [0]{\begingroup \@sanitize@url \@href}%
\providecommand \@href[1]{\@@startlink{#1}\@@href}%
\providecommand \@@href[1]{\endgroup#1\@@endlink}%
\providecommand \@sanitize@url [0]{\catcode `\\12\catcode `\$12\catcode
  `\&12\catcode `\#12\catcode `\^12\catcode `\_12\catcode `\%12\relax}%
\providecommand \@@startlink[1]{}%
\providecommand \@@endlink[0]{}%
\providecommand \url  [0]{\begingroup\@sanitize@url \@url }%
\providecommand \@url [1]{\endgroup\@href {#1}{\urlprefix }}%
\providecommand \urlprefix  [0]{URL }%
\providecommand \Eprint [0]{\href }%
\providecommand \doibase [0]{http://dx.doi.org/}%
\providecommand \selectlanguage [0]{\@gobble}%
\providecommand \bibinfo  [0]{\@secondoftwo}%
\providecommand \bibfield  [0]{\@secondoftwo}%
\providecommand \translation [1]{[#1]}%
\providecommand \BibitemOpen [0]{}%
\providecommand \bibitemStop [0]{}%
\providecommand \bibitemNoStop [0]{.\EOS\space}%
\providecommand \EOS [0]{\spacefactor3000\relax}%
\providecommand \BibitemShut  [1]{\csname bibitem#1\endcsname}%
\let\auto@bib@innerbib\@empty
\bibitem [{\citenamefont {Lubashevsky}\ \emph {et~al.}(2012)\citenamefont
  {Lubashevsky}, \citenamefont {Lahoud}, \citenamefont {Chashka}, \citenamefont
  {Podolsky},\ and\ \citenamefont {Kanigel}}]{NatPhys.8.309}%
  \BibitemOpen
  \bibfield  {author} {\bibinfo {author} {\bibfnamefont {Y.}~\bibnamefont
  {Lubashevsky}}, \bibinfo {author} {\bibfnamefont {E.}~\bibnamefont {Lahoud}},
  \bibinfo {author} {\bibfnamefont {K.}~\bibnamefont {Chashka}}, \bibinfo
  {author} {\bibfnamefont {D.}~\bibnamefont {Podolsky}}, \ and\ \bibinfo
  {author} {\bibfnamefont {A.}~\bibnamefont {Kanigel}},\ }\href {\doibase
  10.1038/nphys2216} {\bibfield  {journal} {\bibinfo  {journal} {Nat. Phys.}\
  }\textbf {\bibinfo {volume} {8}},\ \bibinfo {pages} {309} (\bibinfo {year}
  {2012})}\BibitemShut {NoStop}%
\bibitem [{\citenamefont {Okazaki}\ \emph {et~al.}(2014)\citenamefont
  {Okazaki}, \citenamefont {Ito}, \citenamefont {Ota}, \citenamefont {Kotani},
  \citenamefont {Shimojima}, \citenamefont {Kiss}, \citenamefont {Watanabe},
  \citenamefont {Chen}, \citenamefont {Niitaka}, \citenamefont {Hanaguri},
  \citenamefont {Takagi}, \citenamefont {Chainani},\ and\ \citenamefont
  {Shin}}]{SciRep.4.4109}%
  \BibitemOpen
  \bibfield  {author} {\bibinfo {author} {\bibfnamefont {K.}~\bibnamefont
  {Okazaki}}, \bibinfo {author} {\bibfnamefont {Y.}~\bibnamefont {Ito}},
  \bibinfo {author} {\bibfnamefont {Y.}~\bibnamefont {Ota}}, \bibinfo {author}
  {\bibfnamefont {Y.}~\bibnamefont {Kotani}}, \bibinfo {author} {\bibfnamefont
  {T.}~\bibnamefont {Shimojima}}, \bibinfo {author} {\bibfnamefont
  {T.}~\bibnamefont {Kiss}}, \bibinfo {author} {\bibfnamefont {S.}~\bibnamefont
  {Watanabe}}, \bibinfo {author} {\bibfnamefont {C.-T.}\ \bibnamefont {Chen}},
  \bibinfo {author} {\bibfnamefont {S.}~\bibnamefont {Niitaka}}, \bibinfo
  {author} {\bibfnamefont {T.}~\bibnamefont {Hanaguri}}, \bibinfo {author}
  {\bibfnamefont {H.}~\bibnamefont {Takagi}}, \bibinfo {author} {\bibfnamefont
  {A.}~\bibnamefont {Chainani}}, \ and\ \bibinfo {author} {\bibfnamefont
  {S.}~\bibnamefont {Shin}},\ }\href {\doibase 10.1038/srep04109} {\bibfield
  {journal} {\bibinfo  {journal} {Sci. Rep.}\ }\textbf {\bibinfo {volume}
  {4}},\ \bibinfo {pages} {4109} (\bibinfo {year} {2014})}\BibitemShut
  {NoStop}%
\bibitem [{\citenamefont {Kasahara}\ \emph {et~al.}(2014)\citenamefont
  {Kasahara}, \citenamefont {Watashige}, \citenamefont {Hanaguri},
  \citenamefont {Kohsaka}, \citenamefont {Yamashita}, \citenamefont
  {Shimoyama}, \citenamefont {Mizukami}, \citenamefont {Endo}, \citenamefont
  {Ikeda}, \citenamefont {Aoyama}, \citenamefont {Terashima}, \citenamefont
  {Uji}, \citenamefont {Wolf}, \citenamefont {von L{\"o}hneysen}, \citenamefont
  {Shibauchi},\ and\ \citenamefont {Matsuda}}]{PNAS.111.16309}%
  \BibitemOpen
  \bibfield  {author} {\bibinfo {author} {\bibfnamefont {S.}~\bibnamefont
  {Kasahara}}, \bibinfo {author} {\bibfnamefont {T.}~\bibnamefont {Watashige}},
  \bibinfo {author} {\bibfnamefont {T.}~\bibnamefont {Hanaguri}}, \bibinfo
  {author} {\bibfnamefont {Y.}~\bibnamefont {Kohsaka}}, \bibinfo {author}
  {\bibfnamefont {T.}~\bibnamefont {Yamashita}}, \bibinfo {author}
  {\bibfnamefont {Y.}~\bibnamefont {Shimoyama}}, \bibinfo {author}
  {\bibfnamefont {Y.}~\bibnamefont {Mizukami}}, \bibinfo {author}
  {\bibfnamefont {R.}~\bibnamefont {Endo}}, \bibinfo {author} {\bibfnamefont
  {H.}~\bibnamefont {Ikeda}}, \bibinfo {author} {\bibfnamefont
  {K.}~\bibnamefont {Aoyama}}, \bibinfo {author} {\bibfnamefont
  {T.}~\bibnamefont {Terashima}}, \bibinfo {author} {\bibfnamefont
  {S.}~\bibnamefont {Uji}}, \bibinfo {author} {\bibfnamefont {T.}~\bibnamefont
  {Wolf}}, \bibinfo {author} {\bibfnamefont {H.}~\bibnamefont {von
  L{\"o}hneysen}}, \bibinfo {author} {\bibfnamefont {T.}~\bibnamefont
  {Shibauchi}}, \ and\ \bibinfo {author} {\bibfnamefont {Y.}~\bibnamefont
  {Matsuda}},\ }\href {\doibase 10.1073/pnas.1413477111} {\bibfield  {journal}
  {\bibinfo  {journal} {Proc. Natl. Acad. Sci. U.S.A.}\ }\textbf {\bibinfo
  {volume} {111}},\ \bibinfo {pages} {16309} (\bibinfo {year}
  {2014})}\BibitemShut {NoStop}%
\bibitem [{\citenamefont {Kasahara}\ \emph {et~al.}(2016)\citenamefont
  {Kasahara}, \citenamefont {Yamashita}, \citenamefont {Shi}, \citenamefont
  {Kobayashi}, \citenamefont {Shimoyama}, \citenamefont {Watashige},
  \citenamefont {Ishida}, \citenamefont {Terashima}, \citenamefont {Wolf},
  \citenamefont {Hardy}, \citenamefont {Meingast}, \citenamefont
  {L{\"o}hneysen}, \citenamefont {Levchenko}, \citenamefont {Shibauchi},\ and\
  \citenamefont {Matsuda}}]{NatCom.7.12843}%
  \BibitemOpen
  \bibfield  {author} {\bibinfo {author} {\bibfnamefont {S.}~\bibnamefont
  {Kasahara}}, \bibinfo {author} {\bibfnamefont {T.}~\bibnamefont {Yamashita}},
  \bibinfo {author} {\bibfnamefont {A.}~\bibnamefont {Shi}}, \bibinfo {author}
  {\bibfnamefont {R.}~\bibnamefont {Kobayashi}}, \bibinfo {author}
  {\bibfnamefont {Y.}~\bibnamefont {Shimoyama}}, \bibinfo {author}
  {\bibfnamefont {T.}~\bibnamefont {Watashige}}, \bibinfo {author}
  {\bibfnamefont {K.}~\bibnamefont {Ishida}}, \bibinfo {author} {\bibfnamefont
  {T.}~\bibnamefont {Terashima}}, \bibinfo {author} {\bibfnamefont
  {T.}~\bibnamefont {Wolf}}, \bibinfo {author} {\bibfnamefont {F.}~\bibnamefont
  {Hardy}}, \bibinfo {author} {\bibfnamefont {C.}~\bibnamefont {Meingast}},
  \bibinfo {author} {\bibfnamefont {H.~v.}\ \bibnamefont {L{\"o}hneysen}},
  \bibinfo {author} {\bibfnamefont {A.}~\bibnamefont {Levchenko}}, \bibinfo
  {author} {\bibfnamefont {T.}~\bibnamefont {Shibauchi}}, \ and\ \bibinfo
  {author} {\bibfnamefont {Y.}~\bibnamefont {Matsuda}},\ }\href
  {http://dx.doi.org/10.1038/ncomms12843} {\bibfield  {journal} {\bibinfo
  {journal} {Nature Communications}\ }\textbf {\bibinfo {volume} {7}},\
  \bibinfo {pages} {12843} (\bibinfo {year} {2016})}\BibitemShut {NoStop}%
\bibitem [{\citenamefont {Kitano}\ \emph {et~al.}(2006)\citenamefont {Kitano},
  \citenamefont {Ohashi}, \citenamefont {Maeda},\ and\ \citenamefont
  {Tsukada}}]{PRB.73.092504}%
  \BibitemOpen
  \bibfield  {author} {\bibinfo {author} {\bibfnamefont {H.}~\bibnamefont
  {Kitano}}, \bibinfo {author} {\bibfnamefont {T.}~\bibnamefont {Ohashi}},
  \bibinfo {author} {\bibfnamefont {A.}~\bibnamefont {Maeda}}, \ and\ \bibinfo
  {author} {\bibfnamefont {I.}~\bibnamefont {Tsukada}},\ }\href {\doibase
  10.1103/PhysRevB.73.092504} {\bibfield  {journal} {\bibinfo  {journal} {Phys.
  Rev. B}\ }\textbf {\bibinfo {volume} {73}},\ \bibinfo {pages} {092504}
  (\bibinfo {year} {2006})}\BibitemShut {NoStop}%
\bibitem [{\citenamefont {Ohashi}\ \emph {et~al.}(2009)\citenamefont {Ohashi},
  \citenamefont {Kitano}, \citenamefont {Tsukada},\ and\ \citenamefont
  {Maeda}}]{Ohashi09}%
  \BibitemOpen
  \bibfield  {author} {\bibinfo {author} {\bibfnamefont {T.}~\bibnamefont
  {Ohashi}}, \bibinfo {author} {\bibfnamefont {H.}~\bibnamefont {Kitano}},
  \bibinfo {author} {\bibfnamefont {I.}~\bibnamefont {Tsukada}}, \ and\
  \bibinfo {author} {\bibfnamefont {A.}~\bibnamefont {Maeda}},\ }\href
  {\doibase 10.1103/PhysRevB.79.184507} {\bibfield  {journal} {\bibinfo
  {journal} {Phys. Rev. B}\ }\textbf {\bibinfo {volume} {79}},\ \bibinfo
  {pages} {184507} (\bibinfo {year} {2009})}\BibitemShut {NoStop}%
\bibitem [{\citenamefont {Chen}\ \emph {et~al.}(2006)\citenamefont {Chen},
  \citenamefont {Levin},\ and\ \citenamefont {Stajic}}]{LTPhys.32.406}%
  \BibitemOpen
  \bibfield  {author} {\bibinfo {author} {\bibfnamefont {Q.}~\bibnamefont
  {Chen}}, \bibinfo {author} {\bibfnamefont {K.}~\bibnamefont {Levin}}, \ and\
  \bibinfo {author} {\bibfnamefont {J.}~\bibnamefont {Stajic}},\ }\href
  {\doibase http://dx.doi.org/10.1063/1.2199443} {\bibfield  {journal}
  {\bibinfo  {journal} {Low Temp. Phys.}\ }\textbf {\bibinfo {volume} {32}},\
  \bibinfo {pages} {406} (\bibinfo {year} {2006})}\BibitemShut {NoStop}%
\bibitem [{\citenamefont {Perali}\ \emph {et~al.}(2002)\citenamefont {Perali},
  \citenamefont {Pieri}, \citenamefont {Strinati},\ and\ \citenamefont
  {Castellani}}]{PhysRevB.66.024510}%
  \BibitemOpen
  \bibfield  {author} {\bibinfo {author} {\bibfnamefont {A.}~\bibnamefont
  {Perali}}, \bibinfo {author} {\bibfnamefont {P.}~\bibnamefont {Pieri}},
  \bibinfo {author} {\bibfnamefont {G.~C.}\ \bibnamefont {Strinati}}, \ and\
  \bibinfo {author} {\bibfnamefont {C.}~\bibnamefont {Castellani}},\ }\href
  {\doibase 10.1103/PhysRevB.66.024510} {\bibfield  {journal} {\bibinfo
  {journal} {Phys. Rev. B}\ }\textbf {\bibinfo {volume} {66}},\ \bibinfo
  {pages} {024510} (\bibinfo {year} {2002})}\BibitemShut {NoStop}%
\bibitem [{\citenamefont {Ranninger}\ and\ \citenamefont
  {Robin}(1996)}]{PhysRevB.53.R11961}%
  \BibitemOpen
  \bibfield  {author} {\bibinfo {author} {\bibfnamefont {J.}~\bibnamefont
  {Ranninger}}\ and\ \bibinfo {author} {\bibfnamefont {J.~M.}\ \bibnamefont
  {Robin}},\ }\href {\doibase 10.1103/PhysRevB.53.R11961} {\bibfield  {journal}
  {\bibinfo  {journal} {Phys. Rev. B}\ }\textbf {\bibinfo {volume} {53}},\
  \bibinfo {pages} {R11961} (\bibinfo {year} {1996})}\BibitemShut {NoStop}%
\bibitem [{\citenamefont {Xu}\ \emph {et~al.}(2000)\citenamefont {Xu},
  \citenamefont {Ong}, \citenamefont {Wang}, \citenamefont {Kakeshita},\ and\
  \citenamefont {Uchida}}]{Nature.406.486}%
  \BibitemOpen
  \bibfield  {author} {\bibinfo {author} {\bibfnamefont {Z.~A.}\ \bibnamefont
  {Xu}}, \bibinfo {author} {\bibfnamefont {N.~P.}\ \bibnamefont {Ong}},
  \bibinfo {author} {\bibfnamefont {Y.}~\bibnamefont {Wang}}, \bibinfo {author}
  {\bibfnamefont {T.}~\bibnamefont {Kakeshita}}, \ and\ \bibinfo {author}
  {\bibfnamefont {S.}~\bibnamefont {Uchida}},\ }\href
  {http://dx.doi.org/10.1038/35020016} {\bibfield  {journal} {\bibinfo
  {journal} {Nature}\ }\textbf {\bibinfo {volume} {406}},\ \bibinfo {pages}
  {486} (\bibinfo {year} {2000})}\BibitemShut {NoStop}%
\bibitem [{\citenamefont {Wang}\ \emph {et~al.}(2001)\citenamefont {Wang},
  \citenamefont {Xu}, \citenamefont {Kakeshita}, \citenamefont {Uchida},
  \citenamefont {Ono}, \citenamefont {Ando},\ and\ \citenamefont
  {Ong}}]{PRB.64.224519}%
  \BibitemOpen
  \bibfield  {author} {\bibinfo {author} {\bibfnamefont {Y.}~\bibnamefont
  {Wang}}, \bibinfo {author} {\bibfnamefont {Z.~A.}\ \bibnamefont {Xu}},
  \bibinfo {author} {\bibfnamefont {T.}~\bibnamefont {Kakeshita}}, \bibinfo
  {author} {\bibfnamefont {S.}~\bibnamefont {Uchida}}, \bibinfo {author}
  {\bibfnamefont {S.}~\bibnamefont {Ono}}, \bibinfo {author} {\bibfnamefont
  {Y.}~\bibnamefont {Ando}}, \ and\ \bibinfo {author} {\bibfnamefont {N.~P.}\
  \bibnamefont {Ong}},\ }\href {\doibase 10.1103/PhysRevB.64.224519} {\bibfield
   {journal} {\bibinfo  {journal} {Phys. Rev. B}\ }\textbf {\bibinfo {volume}
  {64}},\ \bibinfo {pages} {224519} (\bibinfo {year} {2001})}\BibitemShut
  {NoStop}%
\bibitem [{\citenamefont {Wang}\ \emph {et~al.}(2006)\citenamefont {Wang},
  \citenamefont {Li},\ and\ \citenamefont {Ong}}]{PRB.73.024510}%
  \BibitemOpen
  \bibfield  {author} {\bibinfo {author} {\bibfnamefont {Y.}~\bibnamefont
  {Wang}}, \bibinfo {author} {\bibfnamefont {L.}~\bibnamefont {Li}}, \ and\
  \bibinfo {author} {\bibfnamefont {N.~P.}\ \bibnamefont {Ong}},\ }\href
  {\doibase 10.1103/PhysRevB.73.024510} {\bibfield  {journal} {\bibinfo
  {journal} {Phys. Rev. B}\ }\textbf {\bibinfo {volume} {73}},\ \bibinfo
  {pages} {024510} (\bibinfo {year} {2006})}\BibitemShut {NoStop}%
\bibitem [{\citenamefont {Wang}\ \emph {et~al.}(2005)\citenamefont {Wang},
  \citenamefont {Li}, \citenamefont {Naughton}, \citenamefont {Gu},
  \citenamefont {Uchida},\ and\ \citenamefont {Ong}}]{PRL.95.247002}%
  \BibitemOpen
  \bibfield  {author} {\bibinfo {author} {\bibfnamefont {Y.}~\bibnamefont
  {Wang}}, \bibinfo {author} {\bibfnamefont {L.}~\bibnamefont {Li}}, \bibinfo
  {author} {\bibfnamefont {M.~J.}\ \bibnamefont {Naughton}}, \bibinfo {author}
  {\bibfnamefont {G.~D.}\ \bibnamefont {Gu}}, \bibinfo {author} {\bibfnamefont
  {S.}~\bibnamefont {Uchida}}, \ and\ \bibinfo {author} {\bibfnamefont {N.~P.}\
  \bibnamefont {Ong}},\ }\href {\doibase 10.1103/PhysRevLett.95.247002}
  {\bibfield  {journal} {\bibinfo  {journal} {Phys. Rev. Lett.}\ }\textbf
  {\bibinfo {volume} {95}},\ \bibinfo {pages} {247002} (\bibinfo {year}
  {2005})}\BibitemShut {NoStop}%
\bibitem [{\citenamefont {Li}\ \emph {et~al.}(2010)\citenamefont {Li},
  \citenamefont {Wang}, \citenamefont {Komiya}, \citenamefont {Ono},
  \citenamefont {Ando}, \citenamefont {Gu},\ and\ \citenamefont
  {Ong}}]{PRB.81.054510}%
  \BibitemOpen
  \bibfield  {author} {\bibinfo {author} {\bibfnamefont {L.}~\bibnamefont
  {Li}}, \bibinfo {author} {\bibfnamefont {Y.}~\bibnamefont {Wang}}, \bibinfo
  {author} {\bibfnamefont {S.}~\bibnamefont {Komiya}}, \bibinfo {author}
  {\bibfnamefont {S.}~\bibnamefont {Ono}}, \bibinfo {author} {\bibfnamefont
  {Y.}~\bibnamefont {Ando}}, \bibinfo {author} {\bibfnamefont {G.~D.}\
  \bibnamefont {Gu}}, \ and\ \bibinfo {author} {\bibfnamefont {N.~P.}\
  \bibnamefont {Ong}},\ }\href {\doibase 10.1103/PhysRevB.81.054510} {\bibfield
   {journal} {\bibinfo  {journal} {Phys. Rev. B}\ }\textbf {\bibinfo {volume}
  {81}},\ \bibinfo {pages} {054510} (\bibinfo {year} {2010})}\BibitemShut
  {NoStop}%
\bibitem [{\citenamefont {Gebre}\ \emph {et~al.}(2011)\citenamefont {Gebre},
  \citenamefont {Li}, \citenamefont {Whalen}, \citenamefont {Conner},
  \citenamefont {Zhou}, \citenamefont {Grissonnanche}, \citenamefont {Kostov},
  \citenamefont {Gurevich}, \citenamefont {Siegrist},\ and\ \citenamefont
  {Balicas}}]{PRB.84.174517}%
  \BibitemOpen
  \bibfield  {author} {\bibinfo {author} {\bibfnamefont {T.}~\bibnamefont
  {Gebre}}, \bibinfo {author} {\bibfnamefont {G.}~\bibnamefont {Li}}, \bibinfo
  {author} {\bibfnamefont {J.~B.}\ \bibnamefont {Whalen}}, \bibinfo {author}
  {\bibfnamefont {B.~S.}\ \bibnamefont {Conner}}, \bibinfo {author}
  {\bibfnamefont {H.~D.}\ \bibnamefont {Zhou}}, \bibinfo {author}
  {\bibfnamefont {G.}~\bibnamefont {Grissonnanche}}, \bibinfo {author}
  {\bibfnamefont {M.~K.}\ \bibnamefont {Kostov}}, \bibinfo {author}
  {\bibfnamefont {A.}~\bibnamefont {Gurevich}}, \bibinfo {author}
  {\bibfnamefont {T.}~\bibnamefont {Siegrist}}, \ and\ \bibinfo {author}
  {\bibfnamefont {L.}~\bibnamefont {Balicas}},\ }\href {\doibase
  10.1103/PhysRevB.84.174517} {\bibfield  {journal} {\bibinfo  {journal} {Phys.
  Rev. B}\ }\textbf {\bibinfo {volume} {84}},\ \bibinfo {pages} {174517}
  (\bibinfo {year} {2011})}\BibitemShut {NoStop}%
\bibitem [{\citenamefont {Schneider}\ \emph {et~al.}(2014)\citenamefont
  {Schneider}, \citenamefont {Zaitsev}, \citenamefont {Fuchs},\ and\
  \citenamefont {von L{\"o}hneysen}}]{JPCM.26.455701}%
  \BibitemOpen
  \bibfield  {author} {\bibinfo {author} {\bibfnamefont {R.}~\bibnamefont
  {Schneider}}, \bibinfo {author} {\bibfnamefont {A.~G.}\ \bibnamefont
  {Zaitsev}}, \bibinfo {author} {\bibfnamefont {D.}~\bibnamefont {Fuchs}}, \
  and\ \bibinfo {author} {\bibfnamefont {H.}~\bibnamefont {von
  L{\"o}hneysen}},\ }\href {http://stacks.iop.org/0953-8984/26/i=45/a=455701}
  {\bibfield  {journal} {\bibinfo  {journal} {Journal of Physics: Condensed
  Matter}\ }\textbf {\bibinfo {volume} {26}},\ \bibinfo {pages} {455701}
  (\bibinfo {year} {2014})}\BibitemShut {NoStop}%
\bibitem [{\citenamefont {Lan}\ \emph {et~al.}(2002)\citenamefont {Lan},
  \citenamefont {Tsai}, \citenamefont {Chang},\ and\ \citenamefont
  {Cheng}}]{SSC.121.575}%
  \BibitemOpen
  \bibfield  {author} {\bibinfo {author} {\bibfnamefont {M.~D.}\ \bibnamefont
  {Lan}}, \bibinfo {author} {\bibfnamefont {P.~L.}\ \bibnamefont {Tsai}},
  \bibinfo {author} {\bibfnamefont {Y.~L.}\ \bibnamefont {Chang}}, \ and\
  \bibinfo {author} {\bibfnamefont {J.~J.}\ \bibnamefont {Cheng}},\ }\href
  {\doibase http://dx.doi.org/10.1016/S0038-1098(02)00042-X} {\bibfield
  {journal} {\bibinfo  {journal} {Solid State Communications}\ }\textbf
  {\bibinfo {volume} {121}},\ \bibinfo {pages} {575} (\bibinfo {year}
  {2002})}\BibitemShut {NoStop}%
\bibitem [{\citenamefont {Dul\ifmmode \check{c}\else
  \v{c}\fi{}i\ifmmode~\acute{c}\else \'{c}\fi{}}\ \emph
  {et~al.}(2003)\citenamefont {Dul\ifmmode \check{c}\else
  \v{c}\fi{}i\ifmmode~\acute{c}\else \'{c}\fi{}}, \citenamefont
  {Po\ifmmode~\check{z}\else \v{z}\fi{}ek}, \citenamefont {Paar}, \citenamefont
  {Choi}, \citenamefont {Kim}, \citenamefont {Kang},\ and\ \citenamefont
  {Lee}}]{PRB.67.020507}%
  \BibitemOpen
  \bibfield  {author} {\bibinfo {author} {\bibfnamefont {A.}~\bibnamefont
  {Dul\ifmmode \check{c}\else \v{c}\fi{}i\ifmmode~\acute{c}\else \'{c}\fi{}}},
  \bibinfo {author} {\bibfnamefont {M.}~\bibnamefont {Po\ifmmode~\check{z}\else
  \v{z}\fi{}ek}}, \bibinfo {author} {\bibfnamefont {D.}~\bibnamefont {Paar}},
  \bibinfo {author} {\bibfnamefont {E.-M.}\ \bibnamefont {Choi}}, \bibinfo
  {author} {\bibfnamefont {H.-J.}\ \bibnamefont {Kim}}, \bibinfo {author}
  {\bibfnamefont {W.~N.}\ \bibnamefont {Kang}}, \ and\ \bibinfo {author}
  {\bibfnamefont {S.-I.}\ \bibnamefont {Lee}},\ }\href {\doibase
  10.1103/PhysRevB.67.020507} {\bibfield  {journal} {\bibinfo  {journal} {Phys.
  Rev. B}\ }\textbf {\bibinfo {volume} {67}},\ \bibinfo {pages} {020507}
  (\bibinfo {year} {2003})}\BibitemShut {NoStop}%
\bibitem [{\citenamefont {Kitano}\ \emph {et~al.}(2008)\citenamefont {Kitano},
  \citenamefont {Ohashi},\ and\ \citenamefont {Maeda}}]{KitanoRSI08}%
  \BibitemOpen
  \bibfield  {author} {\bibinfo {author} {\bibfnamefont {H.}~\bibnamefont
  {Kitano}}, \bibinfo {author} {\bibfnamefont {T.}~\bibnamefont {Ohashi}}, \
  and\ \bibinfo {author} {\bibfnamefont {A.}~\bibnamefont {Maeda}},\ }\href
  {\doibase 10.1063/1.2954957} {\bibfield  {journal} {\bibinfo  {journal}
  {Review of Scientific Instruments}\ }\textbf {\bibinfo {volume} {79}},\
  \bibinfo {pages} {074701} (\bibinfo {year} {2008})}\BibitemShut {NoStop}%
\bibitem [{\citenamefont {Imai}\ \emph
  {et~al.}(2010{\natexlab{a}})\citenamefont {Imai}, \citenamefont {Tanaka},
  \citenamefont {Akiike}, \citenamefont {Hanawa}, \citenamefont {Tsukada},\
  and\ \citenamefont {Maeda}}]{Imai09}%
  \BibitemOpen
  \bibfield  {author} {\bibinfo {author} {\bibfnamefont {Y.}~\bibnamefont
  {Imai}}, \bibinfo {author} {\bibfnamefont {R.}~\bibnamefont {Tanaka}},
  \bibinfo {author} {\bibfnamefont {T.}~\bibnamefont {Akiike}}, \bibinfo
  {author} {\bibfnamefont {M.}~\bibnamefont {Hanawa}}, \bibinfo {author}
  {\bibfnamefont {I.}~\bibnamefont {Tsukada}}, \ and\ \bibinfo {author}
  {\bibfnamefont {A.}~\bibnamefont {Maeda}},\ }\href
  {http://stacks.iop.org/1347-4065/49/i=2R/a=023101} {\bibfield  {journal}
  {\bibinfo  {journal} {Jpn. J. Appl. Phys.}\ }\textbf {\bibinfo {volume}
  {49}},\ \bibinfo {pages} {023101} (\bibinfo {year}
  {2010}{\natexlab{a}})}\BibitemShut {NoStop}%
\bibitem [{\citenamefont {Imai}\ \emph
  {et~al.}(2010{\natexlab{b}})\citenamefont {Imai}, \citenamefont {Akiike},
  \citenamefont {Hanawa}, \citenamefont {Tsukada}, \citenamefont {Ichinose},
  \citenamefont {Maeda}, \citenamefont {Hikage}, \citenamefont {Kawaguchi},\
  and\ \citenamefont {Ikuta}}]{Imai10}%
  \BibitemOpen
  \bibfield  {author} {\bibinfo {author} {\bibfnamefont {Y.}~\bibnamefont
  {Imai}}, \bibinfo {author} {\bibfnamefont {T.}~\bibnamefont {Akiike}},
  \bibinfo {author} {\bibfnamefont {M.}~\bibnamefont {Hanawa}}, \bibinfo
  {author} {\bibfnamefont {I.}~\bibnamefont {Tsukada}}, \bibinfo {author}
  {\bibfnamefont {A.}~\bibnamefont {Ichinose}}, \bibinfo {author}
  {\bibfnamefont {A.}~\bibnamefont {Maeda}}, \bibinfo {author} {\bibfnamefont
  {T.}~\bibnamefont {Hikage}}, \bibinfo {author} {\bibfnamefont
  {T.}~\bibnamefont {Kawaguchi}}, \ and\ \bibinfo {author} {\bibfnamefont
  {H.}~\bibnamefont {Ikuta}},\ }\href
  {http://stacks.iop.org/1882-0786/3/i=4/a=043102} {\bibfield  {journal}
  {\bibinfo  {journal} {Appl. Phys. Express}\ }\textbf {\bibinfo {volume}
  {3}},\ \bibinfo {pages} {043102} (\bibinfo {year}
  {2010}{\natexlab{b}})}\BibitemShut {NoStop}%
\bibitem [{\citenamefont {Tsukada}\ \emph {et~al.}(2011)\citenamefont
  {Tsukada}, \citenamefont {Hanawa}, \citenamefont {Akiike}, \citenamefont
  {Nabeshima}, \citenamefont {Imai}, \citenamefont {Ichinose}, \citenamefont
  {Komiya}, \citenamefont {Hikage}, \citenamefont {Kawaguchi}, \citenamefont
  {Ikuta},\ and\ \citenamefont {Maeda}}]{Tsukada11}%
  \BibitemOpen
  \bibfield  {author} {\bibinfo {author} {\bibfnamefont {I.}~\bibnamefont
  {Tsukada}}, \bibinfo {author} {\bibfnamefont {M.}~\bibnamefont {Hanawa}},
  \bibinfo {author} {\bibfnamefont {T.}~\bibnamefont {Akiike}}, \bibinfo
  {author} {\bibfnamefont {F.}~\bibnamefont {Nabeshima}}, \bibinfo {author}
  {\bibfnamefont {Y.}~\bibnamefont {Imai}}, \bibinfo {author} {\bibfnamefont
  {A.}~\bibnamefont {Ichinose}}, \bibinfo {author} {\bibfnamefont
  {S.}~\bibnamefont {Komiya}}, \bibinfo {author} {\bibfnamefont
  {T.}~\bibnamefont {Hikage}}, \bibinfo {author} {\bibfnamefont
  {T.}~\bibnamefont {Kawaguchi}}, \bibinfo {author} {\bibfnamefont
  {H.}~\bibnamefont {Ikuta}}, \ and\ \bibinfo {author} {\bibfnamefont
  {A.}~\bibnamefont {Maeda}},\ }\href
  {http://stacks.iop.org/1882-0786/4/i=5/a=053101} {\bibfield  {journal}
  {\bibinfo  {journal} {Appl. Phys. Express}\ }\textbf {\bibinfo {volume}
  {4}},\ \bibinfo {pages} {053101} (\bibinfo {year} {2011})}\BibitemShut
  {NoStop}%
\bibitem [{\citenamefont {Nabeshima}\ \emph {et~al.}(2013)\citenamefont
  {Nabeshima}, \citenamefont {Imai}, \citenamefont {Hanawa}, \citenamefont
  {Tsukada},\ and\ \citenamefont {Maeda}}]{Nabe13}%
  \BibitemOpen
  \bibfield  {author} {\bibinfo {author} {\bibfnamefont {F.}~\bibnamefont
  {Nabeshima}}, \bibinfo {author} {\bibfnamefont {Y.}~\bibnamefont {Imai}},
  \bibinfo {author} {\bibfnamefont {M.}~\bibnamefont {Hanawa}}, \bibinfo
  {author} {\bibfnamefont {I.}~\bibnamefont {Tsukada}}, \ and\ \bibinfo
  {author} {\bibfnamefont {A.}~\bibnamefont {Maeda}},\ }\href {\doibase
  10.1063/1.4826945} {\bibfield  {journal} {\bibinfo  {journal} {Appl. Phys.
  Lett.}\ }\textbf {\bibinfo {volume} {103}},\ \bibinfo {pages} {172602}
  (\bibinfo {year} {2013})}\BibitemShut {NoStop}%
\bibitem [{\citenamefont {Steinberg}\ \emph {et~al.}(2008)\citenamefont
  {Steinberg}, \citenamefont {Scheffler},\ and\ \citenamefont
  {Dressel}}]{PRB77.214517}%
  \BibitemOpen
  \bibfield  {author} {\bibinfo {author} {\bibfnamefont {K.}~\bibnamefont
  {Steinberg}}, \bibinfo {author} {\bibfnamefont {M.}~\bibnamefont
  {Scheffler}}, \ and\ \bibinfo {author} {\bibfnamefont {M.}~\bibnamefont
  {Dressel}},\ }\href {\doibase 10.1103/PhysRevB.77.214517} {\bibfield
  {journal} {\bibinfo  {journal} {Phys. Rev. B}\ }\textbf {\bibinfo {volume}
  {77}},\ \bibinfo {pages} {214517} (\bibinfo {year} {2008})}\BibitemShut
  {NoStop}%
\bibitem [{\citenamefont {Liu}\ \emph {et~al.}(2011)\citenamefont {Liu},
  \citenamefont {Kim}, \citenamefont {Sambandamurthy},\ and\ \citenamefont
  {Armitage}}]{PRB84.024511}%
  \BibitemOpen
  \bibfield  {author} {\bibinfo {author} {\bibfnamefont {W.}~\bibnamefont
  {Liu}}, \bibinfo {author} {\bibfnamefont {M.}~\bibnamefont {Kim}}, \bibinfo
  {author} {\bibfnamefont {G.}~\bibnamefont {Sambandamurthy}}, \ and\ \bibinfo
  {author} {\bibfnamefont {N.~P.}\ \bibnamefont {Armitage}},\ }\href {\doibase
  10.1103/PhysRevB.84.024511} {\bibfield  {journal} {\bibinfo  {journal} {Phys.
  Rev. B}\ }\textbf {\bibinfo {volume} {84}},\ \bibinfo {pages} {024511}
  (\bibinfo {year} {2011})}\BibitemShut {NoStop}%
\bibitem [{\citenamefont {Fisher}\ \emph {et~al.}(1991)\citenamefont {Fisher},
  \citenamefont {Fisher},\ and\ \citenamefont {Huse}}]{PRB.43.130}%
  \BibitemOpen
  \bibfield  {author} {\bibinfo {author} {\bibfnamefont {D.~S.}\ \bibnamefont
  {Fisher}}, \bibinfo {author} {\bibfnamefont {M.~P.~A.}\ \bibnamefont
  {Fisher}}, \ and\ \bibinfo {author} {\bibfnamefont {D.~A.}\ \bibnamefont
  {Huse}},\ }\href {\doibase 10.1103/PhysRevB.43.130} {\bibfield  {journal}
  {\bibinfo  {journal} {Phys. Rev. B}\ }\textbf {\bibinfo {volume} {43}},\
  \bibinfo {pages} {130} (\bibinfo {year} {1991})}\BibitemShut {NoStop}%
\bibitem [{\citenamefont {Sawada}\ \emph {et~al.}(2016)\citenamefont {Sawada},
  \citenamefont {Nabeshima}, \citenamefont {Imai},\ and\ \citenamefont
  {Maeda}}]{ys16jpsj}%
  \BibitemOpen
  \bibfield  {author} {\bibinfo {author} {\bibfnamefont {Y.}~\bibnamefont
  {Sawada}}, \bibinfo {author} {\bibfnamefont {F.}~\bibnamefont {Nabeshima}},
  \bibinfo {author} {\bibfnamefont {Y.}~\bibnamefont {Imai}}, \ and\ \bibinfo
  {author} {\bibfnamefont {A.}~\bibnamefont {Maeda}},\ }\href {\doibase
  http://dx.doi.org/10.7566/JPSJ.85.073703} {\bibfield  {journal} {\bibinfo
  {journal} {J. Phys. Soc. Jpn.}\ }\textbf {\bibinfo {volume} {85}},\ \bibinfo
  {pages} {073703} (\bibinfo {year} {2016})}\BibitemShut {NoStop}%
\bibitem [{Note1()}]{Note1}%
  \BibitemOpen
  \bibinfo {note} {According to ref.\cite {ys16jpsj}, $\xi _\perp \sim 1$ nm in
  the zero temperature limit. Then, we obtain $\xi (T) = \xi (0~\protect
  \mathrm K) \times (T/T_{\protect \mathrm c} - 1)^{-\nu } \sim 100$ nm ($\nu =
  0.67$ for the 3D-XY model) at temperature of $(T/T_{\protect \mathrm c} - 1)
  \sim 10^{-3}$. This is consistent with our results of the scaling analysis
  shown in Fig. \ref {G-Scaling}, that the 30-nm and 60-nm samples were 2D and
  the 116-nm and 150-nm samples were 3D at the temperature.}\BibitemShut
  {Stop}%
\bibitem [{\citenamefont {Ohashi}\ \emph {et~al.}(2006)\citenamefont {Ohashi},
  \citenamefont {Kitano}, \citenamefont {Maeda}, \citenamefont {Akaike},\ and\
  \citenamefont {Fujimaki}}]{PRB.73.174522}%
  \BibitemOpen
  \bibfield  {author} {\bibinfo {author} {\bibfnamefont {T.}~\bibnamefont
  {Ohashi}}, \bibinfo {author} {\bibfnamefont {H.}~\bibnamefont {Kitano}},
  \bibinfo {author} {\bibfnamefont {A.}~\bibnamefont {Maeda}}, \bibinfo
  {author} {\bibfnamefont {H.}~\bibnamefont {Akaike}}, \ and\ \bibinfo {author}
  {\bibfnamefont {A.}~\bibnamefont {Fujimaki}},\ }\href {\doibase
  10.1103/PhysRevB.73.174522} {\bibfield  {journal} {\bibinfo  {journal} {Phys.
  Rev. B}\ }\textbf {\bibinfo {volume} {73}},\ \bibinfo {pages} {174522}
  (\bibinfo {year} {2006})}\BibitemShut {NoStop}%
\bibitem [{\citenamefont {Berezinskii}(1970)}]{JETP.32.493}%
  \BibitemOpen
  \bibfield  {author} {\bibinfo {author} {\bibfnamefont {V.~L.}\ \bibnamefont
  {Berezinskii}},\ }\href@noop {} {\bibfield  {journal} {\bibinfo  {journal}
  {Sov. Phys. JETP}\ }\textbf {\bibinfo {volume} {32}},\ \bibinfo {pages} {493}
  (\bibinfo {year} {1970})}\BibitemShut {NoStop}%
\bibitem [{\citenamefont {Kosterlitz}\ and\ \citenamefont
  {Thouless}(1973)}]{KT1973}%
  \BibitemOpen
  \bibfield  {author} {\bibinfo {author} {\bibfnamefont {J.~M.}\ \bibnamefont
  {Kosterlitz}}\ and\ \bibinfo {author} {\bibfnamefont {D.~J.}\ \bibnamefont
  {Thouless}},\ }\href {http://stacks.iop.org/0022-3719/6/i=7/a=010} {\bibfield
   {journal} {\bibinfo  {journal} {Journal of Physics C: Solid State Physics}\
  }\textbf {\bibinfo {volume} {6}},\ \bibinfo {pages} {1181} (\bibinfo {year}
  {1973})}\BibitemShut {NoStop}%
\bibitem [{\citenamefont {Imai}\ \emph {et~al.}(2015)\citenamefont {Imai},
  \citenamefont {Sawada}, \citenamefont {Nabeshima},\ and\ \citenamefont
  {Maeda}}]{yi15pnas}%
  \BibitemOpen
  \bibfield  {author} {\bibinfo {author} {\bibfnamefont {Y.}~\bibnamefont
  {Imai}}, \bibinfo {author} {\bibfnamefont {Y.}~\bibnamefont {Sawada}},
  \bibinfo {author} {\bibfnamefont {F.}~\bibnamefont {Nabeshima}}, \ and\
  \bibinfo {author} {\bibfnamefont {A.}~\bibnamefont {Maeda}},\ }\href
  {\doibase 10.1073/pnas.1418994112} {\bibfield  {journal} {\bibinfo  {journal}
  {Proc. Natl. Acad. Sci. U.S.A.}\ }\textbf {\bibinfo {volume} {112}},\
  \bibinfo {pages} {1937} (\bibinfo {year} {2015})}\BibitemShut {NoStop}%
\bibitem [{\citenamefont {Imai}\ \emph {et~al.}(2017)\citenamefont {Imai},
  \citenamefont {Sawada}, \citenamefont {Nabeshima}, \citenamefont {Asami},
  \citenamefont {Kawai},\ and\ \citenamefont {Maeda}}]{yiSciRep17}%
  \BibitemOpen
  \bibfield  {author} {\bibinfo {author} {\bibfnamefont {Y.}~\bibnamefont
  {Imai}}, \bibinfo {author} {\bibfnamefont {Y.}~\bibnamefont {Sawada}},
  \bibinfo {author} {\bibfnamefont {F.}~\bibnamefont {Nabeshima}}, \bibinfo
  {author} {\bibfnamefont {D.}~\bibnamefont {Asami}}, \bibinfo {author}
  {\bibfnamefont {M.}~\bibnamefont {Kawai}}, \ and\ \bibinfo {author}
  {\bibfnamefont {A.}~\bibnamefont {Maeda}},\ }\href
  {http://dx.doi.org/10.1038/srep46653} {\bibfield  {journal} {\bibinfo
  {journal} {Sci. Rep.}\ }\textbf {\bibinfo {volume} {7}},\ \bibinfo {pages}
  {46653} (\bibinfo {year} {2017})}\BibitemShut {NoStop}%
\bibitem [{Note2()}]{Note2}%
  \BibitemOpen
  \bibinfo {note} {We took the temperature-dependent $S_{11}$ of an FeTe film
  as load data. In addition, to reduce errors in arg($S_{11}$)
  between the $sample$ and the $reference$ measurements, which is due to 
  the possible difference in the strength of
  contact between the film and the coaxial cable, we added a constant value to
  arg($S_{11}$) of the load reference so that arg($S_{11}^{\protect \mathrm {load}}$)
  at 50 K to be that of superconducting
  sample at the same temperature. Then, we calibrate $S_{11}$ of the sample
  using these data as well as those of Au (Short) and Teflon
  (Open).}\BibitemShut {Stop}%
\bibitem [{\citenamefont {Nakamura}\ \emph {et~al.}(2012)\citenamefont
  {Nakamura}, \citenamefont {Imai}, \citenamefont {Maeda},\ and\ \citenamefont
  {Tsukada}}]{Nakamura12}%
  \BibitemOpen
  \bibfield  {author} {\bibinfo {author} {\bibfnamefont {D.}~\bibnamefont
  {Nakamura}}, \bibinfo {author} {\bibfnamefont {Y.}~\bibnamefont {Imai}},
  \bibinfo {author} {\bibfnamefont {A.}~\bibnamefont {Maeda}}, \ and\ \bibinfo
  {author} {\bibfnamefont {I.}~\bibnamefont {Tsukada}},\ }\href {\doibase
  10.1143/JPSJ.81.044709} {\bibfield  {journal} {\bibinfo  {journal} {Journal
  of the Physical Society of Japan}\ }\textbf {\bibinfo {volume} {81}},\
  \bibinfo {pages} {044709} (\bibinfo {year} {2012})}\BibitemShut {NoStop}%
\bibitem [{\citenamefont {Schmidt}(1968)}]{Schmidt68}%
  \BibitemOpen
  \bibfield  {author} {\bibinfo {author} {\bibfnamefont {H.}~\bibnamefont
  {Schmidt}},\ }\href@noop {} {\bibfield  {journal} {\bibinfo  {journal} {Z.
  Phys.}\ }\textbf {\bibinfo {volume} {216}},\ \bibinfo {pages} {336 }
  (\bibinfo {year} {1968})}\BibitemShut {NoStop}%
\end{thebibliography}
%

\end{document}